\documentclass[preprint,showpacs,amssymb,aps,prd]{revtex4}

\begin{document}

%
\title{Renormalization in conformal quantum mechanics}

\author{ Horacio E. Camblong$^{a}$\footnotetext{{\em E-mail
addresses\/}: camblong@usfca.edu
(H.E. Camblong), ordonez@uh.edu (C.R. Ord\'{o}\~{n}ez)} 
and
Carlos R. Ord\'{o}\~{n}ez$^{b,c}$}

\affiliation{
$^{a}$
Department of Physics, University of San Francisco, San
Francisco, California 94117-1080, USA
\\
$^{b}$ Department of Physics, University of Houston, Houston,
Texas 77204-5506, USA
\\
$^{c}$
World Laboratory Center for Pan-American Collaboration in Science and
Technology,
\\
University of Houston Center, Houston, Texas 77204-5506, USA
}

\begin{abstract}
The singular behavior of conformal interactions is examined
within a comparative analysis of renormalization frameworks.
The {\em effective\/} approach---inspired by the effective-field theory
program---and its connection with the
core framework are highlighted. 
Applications include black-hole thermodynamics, 
molecular dipole-bound anions,
the Efimov effect, and various regimes of QED.

\end{abstract}


\maketitle

\section{\large\bf I\lowercase{ntroduction}}
\label{sec:introduction}

Conformal quantum mechanics is based upon the class of Lagrangians whose 
action is invariant under translations, dilations, and special conformal 
transformations in time~\cite{jackiw:72,alfaro:76}.
The interaction potential $V({\bf r})$, associated with the Hamiltonian
\begin{equation}
H
= \frac{ p^{2}}{2m}  + V({\bf r})
\;  ,
\label{eq:CQM_Hamiltonian}
\end{equation}
is homogeneous 
of degree $-2$~\cite{camblong:dt} and generates the 
SO(2,1) {\em conformal algebra\/}~\cite{jackiw:80_90}
\begin{equation}
[D,H]
= - i \hbar H
  \;  ,
\; \;
[K,H]
= - 2 i \hbar D
\;  ,
\; \;
[D, K]
=  i \hbar K
\;  ,
\label{eq:naive_commutators}
\end{equation}
which involves
the dilation operator
\begin{displaymath}
D= t H - \frac{  {\bf r} \cdot {\bf p} + {\bf p} \cdot
{\bf r} }{4}
\end{displaymath}
and the special conformal operator
\begin{displaymath}
K= t^{2} H -  \frac{t \, ({\bf p} \cdot {\bf r} + {\bf r} \cdot
{\bf p}) }{2}
 +  \frac{m
r^{2} }{2}
\; .
\end{displaymath}
Systems of this kind are singular and 
ill-defined for a sufficiently {\em strong coupling\/}, as a 
corollary of the existence of scale 
symmetry~\cite{camblong:dt} and
no further quantum-mechanical analysis appears to be acceptable.
However, an alternative viewpoint is possible:
the singular behavior of the conformal interaction
reveals the existence of additional {\em ultraviolet physics\/},
just like in quantum field theory. This {\em renormalization
interpretation\/}, first discussed for the two-dimensional contact 
interaction~\cite{jackiw:91-beg,hua:92,tarrach},
was extended to higher dimensions and to other
interactions~\cite{adhikari:95_97}.
In this procedure, detailed knowledge of the short-distances 
physics is not needed but it is possible to answer 
well-posed questions for the original problem
by the use of regularization and renormalization.
In essence, these tools permit a consistent and physically transparent
analysis of {\em singular quantum mechanics\/}.

In this Letter, we consider the long-range
representative of conformal
quantum mechanics in its spherically symmetric form.
This interaction is considerably
more pathological than the contact potential and has subtler
renormalization properties,
which we investigate by:
(i) developing the renormalization frameworks within a
{\em unified approach\/},
with a real-space regularization procedure;
(ii) highlighting a novel effective framework that captures
the {\em universal predictions\/} for 
fixed long-distance physics;
(iii) illustrating this generic toolbox for 
several {\em physical realizations\/}.

\section{\large\bf R\lowercase{eal-Space Regularization}}
\label{real_space_regularization}

Our Letter is based on a renormalization procedure
in real space for singular interactions, 
within the philosophy of the {\em effective-field theory 
program\/}~\cite{EFT}.
This is justified by the 
effective nature of any physical description of a system for which 
the ultraviolet physics is replaced over length scales $r \alt a$.
Correspondingly, the ensuing 
effective theory has predictability in 
the realm of energies of magnitude 
$|E| \ll {\mathcal E}_{a} 
\equiv \hbar^{2}/2m a^{2} $, defining
a conformal domain (possibly limited by the onset 
of infrared physics).

The real-space regularization procedure involves a regulator $a$ and 
a ``regularizing core'' $V^{(<)}({\bf r})  $
that, over scales $ r \alt a$, replaces the 
singular potential, while its functional form is only retained in  
the exterior region~\cite{landau:77}:
$V({\bf r}) \equiv V^{(>)}({\bf r})  $.
The regularizing core can be parametrized 
with a dimensionless function  ${\mathcal F} ({\bf r})$, in the form
\begin{equation}
V^{(<)}({\bf r}) 
= \frac{\hbar^{2}}{2m}  
\, {\mathcal V^{(<)}} ({\bf r})
= - \frac{\hbar^{2}}{2m} \, \frac{\aleph}{a^{2}} 
\, {\mathcal F} ({\bf r})
\; ,
\label{eq:regularizing_core}
\end{equation}
where ${\mathcal F} ({\bf r}) \geq 0 $ is normalized with
\begin{displaymath}
{\rm max}_{ {\bf r} \in [0,a]} 
\left[ {\mathcal F} ({\bf r}) \right]= 1
\; , 
\end{displaymath}
so that $\aleph$ measures its dimensionless depth.
In the exterior region,
the singular conformal potential 
$V({\bf r}) 
= - 
g/r^{2} 
$
yields 
\begin{equation}
V^{(>)}({\bf r}) 
= \frac{\hbar^{2}}{2m}  
\, {\mathcal V^{(>)}} ({\bf r})
= - \frac{\hbar^{2}}{2m} \, \frac{\lambda}{r^{2}} 
\; 
\label{eq:external_ISP}
\end{equation}
(where $\lambda = 2m g/\hbar^{2}$).
Moreover, this treatment can 
be generalized to the 
anisotropic inverse square potential~\cite{cam:anomaly_qm_ISP_anyD},
e.g.,
in  molecular physics~\cite{molecular_dipole_anomaly}.

As a first step,
we consider the $d$-dimensional effective radial Schr\"odinger equation 
\begin{equation}
\left[
\frac{d^{2}}{dr^{2}}
+
k^{2} - \frac{(l+\nu)^{2}- 1/4}{ r^{2} } -
 {\mathcal V} (r)
\right] u (r)
= 0
\; ,
\label{eq:reduced_Schrodinger_u}
\end{equation}
where 
$\nu 
 = d/2 -1
$
 and
$
\Psi ({\bf r})= 
Y_{l m}( {\bf \Omega})  \,  u (r) /r^{\nu+1/2 } $
is the full-fledged separable solution, in which
$Y_{l m }( {\bf \Omega})  $ stands for the ultraspherical 
harmonics on $S^{d-1}$~\cite{ultraspherical_harmonics}.

For the bound-state sector, the criterion of square integrability 
provides the solution 
\begin{equation}
 u^{(>)}(r) 
 = 
A_{l,\nu} \,
\sqrt{r}
\, 
K_{i \Theta } (\kappa r)
\label{eq:ISP_BS_wf_>}
\; ,
\end{equation} 
for  $r>a$ and energy $E=  -\hbar^{2} \kappa^2/2m<0$,
in terms of
the Macdonald function~\cite{macdonald_function} $K_{i \Theta } (z)$
of imaginary order
defined from the {\em conformal coupling\/} $\lambda$ through
a {\em conformal parameter\/}
\begin{equation}
 \Theta 
=
\sqrt{ \lambda - (l+\nu)^{2}}
\; . 
\label{eq:Theta_coupling}
\end{equation}
In particular,  Eq.~(\ref{eq:Theta_coupling}) 
leads to the definition of the critical coupling
\begin{equation}
\lambda^{(*)} 
= \left(
l + \nu
\right)^{2}
\; .
\label{eq:critical_coupling}
\end{equation}
For the strong-coupling
regime, $ \lambda \geq \lambda^{(*)} $,
Eq.~(\ref{eq:ISP_BS_wf_>})
gives the oscillatory behavior~\cite{cas:50,mor:53}
\begin{equation}
 u^{(>)}(r) 
  \stackrel{(a< r \sim 0)}{=}
- A_{l,\nu} \;
\sqrt{r}
\,
\sqrt{ \frac{\pi}{ \Theta 
\sinh  \left( \pi \Theta \right) } }
\,
\sin
\left\{
\Theta
\left[
 \ln \left( \frac{ \kappa r }{2} \right) 
+ \gamma_{\Theta} 
\right]
\right\}
\,
\left[ 1 + O \left( [\kappa r ]^{2} \right)  
\right] 
\;  
\label{eq:MacDonald_asymptotic}
\end{equation}
near the origin, 
where
\begin{displaymath}
\gamma_{\Theta} = - 
\left\{
{\rm phase} 
\left[
\Gamma (1 +i\Theta )
\right]
\right\}/\Theta
\end{displaymath}
generalizes the Euler-Mascheroni constant,  
to which it reduces in the limit $\Theta \rightarrow 0$. 

Furthermore, when a regularizing core is considered,
the interior solution (for $r<a$)
 \begin{equation}
u^{(<)}(r)= B_{l,\nu} \, \sqrt{r}
\,
w_{l+\nu}( \tilde{k} r; \tilde{k})
\;
\label{eq:ISP_BS_wf_<}
\end{equation} 
satisfies the regular boundary condition at the origin;
here, $\tilde{k}$  is implicitly defined from
\begin{equation}
(\tilde{k} a )^2
+( \kappa a)^{2}  
= \aleph 
\; ,
\label{eq:pythagorean}
\end{equation}
while
${\mathcal F} ({\bf r})$ and $w_{l+\nu}$
may depend upon additional parameters.
The central equation is
\begin{equation}
{\cal L}^{(<)} (\tilde{k} a; \tilde{k} )
=
{\cal L}^{(>)} (\kappa a)
\;  ,
\label{eq:log_derivatives}
\end{equation}
which describes, for
the reduced wave functions 
\begin{equation}
v (r) = \frac{ u (r)}{ \sqrt{r} }
=
\left\{
\begin{array}{ll}
v^{(<)} (r)
\propto
w_{l+\nu}( \tilde{k} r; \tilde{k})
& {\rm for} \; r < a
\\
v^{(>)} (\kappa r)
\propto
 K_{i\Theta} (\kappa r) & {\rm for} \; r> a
\end{array}
\right.
\; ,
\label{reduced_Bessel-like_radial_wf}
\end{equation}
 the continuity 
of the logarithmic derivatives
\begin{equation}
{\cal L} 
\equiv  
 \frac{d \ln v(r)}{d \ln r} 
\; .
\end{equation}
Equation~(\ref{eq:log_derivatives}) is further supplemented 
by the continuity of the wave function,
\begin{equation}
 B_{l,\nu} \, 
w_{l+\nu}( \tilde{k} a; \tilde{k})
= A_{l,\nu} \, K_{i \Theta}  ( \kappa a)
\; .
\end{equation}
From Eqs.~(\ref{eq:MacDonald_asymptotic})
and (\ref{eq:pythagorean}),
the fundamental condition~(\ref{eq:log_derivatives}) reduces to the form
\begin{equation}
\cot \left[
\alpha 
\left( 
\Theta, \kappa a 
\right)
\right]
  \stackrel{( \kappa a \ll 1)}{\sim}
\frac{1}{\Theta}
 \;
{\cal L}^{(<)} \left( {\aleph} \right)
\;  ,
\label{eq:log_derivatives_asymptotic}
\end{equation}
where
\begin{equation}
\alpha \left( \Theta, \kappa a \right)
\equiv
\Theta \left[ \ln \left( \frac{\kappa a}{2} \right) + \gamma_{\Theta} 
\right]
\;  
\label{eq:alpha_definition}
\end{equation}
and, {\em by abuse of notation\/},
we define
${\cal L}^{(<)} \left( {\aleph} \right)$ 
through the limit
\begin{equation}
{\cal L}^{(<)} (\tilde{k} 
a; \tilde{k} )
  \stackrel{( \kappa a \ll 1)}{\sim}
{\cal L}^{(<)} (\sqrt{\aleph}; \tilde{k} )
\equiv
{\cal L}^{(<)} \left( {\aleph} \right)
\;  .
\end{equation}
Equation~(\ref{eq:log_derivatives_asymptotic})
leads to a precise definition of 
renormalization for this system.

\section{\large\bf E\lowercase{ffective renormalization framework}}

In the effective framework,
the regularization of the system is defined in a manner consistent
with two fundamental requirements,
dictated by the physics of relevant realizations:
\begin{enumerate}
\item
The
{\em existence of bound states with finite energy\/},
set by the ultraviolet physics.
\item
The restriction of the
{\em conformal coupling 
to be fixed by the long-distance behavior\/}.
\end{enumerate}

The relative simplicity of the conformal physics is due to
its {\em scaling properties\/}, which permit a direct analysis based on the 
parameter $\kappa$ associated with bound states 
(see Eq.~(\ref{eq:ISP_BS_wf_>}))
and its {\em hierarchical comparison\/}
with other relevant scales.
Then, the condition $\kappa a \ll 1$
can be applied systematically---however,
for $\kappa a \agt 1$,
no prediction can be made as ``new'' ultraviolet physics 
supersedes the conformal interaction.
More precisely, consistent use of the condition $\kappa a \ll 1$,
in Eq.~(\ref{eq:log_derivatives_asymptotic}) and related expressions,
establishes the claim we made at the beginning of the previous section:
the {\em effective theory\/} inherits predictability in the energy realm 
$|E|  \ll {\mathcal E}_{a} $ and generates the
{\em universal behavior\/} of the spherically symmetric long-range 
conformal interaction, 
{\em regardless of the details of the ultraviolet physics\/}.

In the conformally invariant domain, i.e., $\kappa a \ll 1$, 
Eq.~(\ref{eq:log_derivatives_asymptotic}) 
leads to the behavior 
 \begin{equation}
\alpha (\Theta, \kappa a) 
\stackrel{( \kappa a \ll 1)}{\sim} 
- \pi \, f_{n}
=
- \pi \, \left( n  + f_{0}  \right) 
\; ,
\label{eq:alpha_conformal_values}
\end{equation}
 where $n$ is a positive integer, 
$f_{n} =  n  + f_{0}  $,
and $f_{0}$ a dimensionless constant
of order one.
The particular value of $f_{0}$, which is sensitively dependent upon
the details of the ultraviolet physics, is not relevant in the determination 
of the {\em universal conformal properties\/}. This can be seen from
Eqs.~(\ref{eq:alpha_definition})
and (\ref{eq:alpha_conformal_values}), which lead to the 
bound-state energy spectrum
\begin{equation}
E_{ n }
=
E_{0}
\,
\exp \left( - \frac{2 \pi n}{\Theta } \right)
\;  ,
\label{eq:cutoff_BS_regularized_energies_phenomenological}
\end{equation}
which is a geometric sequence with ratio
\begin{equation}
\eta = \exp \left( - \frac{2 \pi }{\Theta } \right)
\; ,
\label{eq:ratio_geometric-sequence}
\end{equation}
starting from
\begin{equation}
E_{0}
=
-
{\mathcal E}_{a} 
\,
(2 e^{- \gamma_{\Theta} })^{2} 
e^{ -2 \pi f_{0}/\Theta}
\; .
\label{eq:GS_energy}
\end{equation}
This universal prediction of conformal quantum mechanics 
could be tested by considering 
\begin{equation}
\epsilon_{n',n} \equiv
\frac{ E_{n'} }{ E_{n} } 
=  
\eta^{ n'-n }
\; ;
\label{eq:ratios_cutoff_BS_regularized_energies_phenomenological}
\end{equation}
these ratios depend on a single physical quantity: the 
{\em conformal parameter\/} $\Theta$ (through the 
exponential~(\ref{eq:ratio_geometric-sequence})).
Even though $f_{0}$ and $a$ still determine the precise value
of $E_{0}$, in this framework, the scale $E_{0}$ 
is either an observable to be adjusted experimentally
or a quantity to be determined from additional ultraviolet-specific 
information. 
In short, the {\em conformal tower\/} of bound 
states~(\ref{eq:cutoff_BS_regularized_energies_phenomenological})
has the following attributes:
\begin{enumerate}

\item 
{\em Universality\/}: it is independent of 
ultraviolet and infrared alterations of the physics,
for the subset of theories with 
{\em conformal coupling set by the long-distance behavior\/}.

\item
{\em Geometric scaling\/}: characterized by the relative
ratios~(\ref{eq:ratios_cutoff_BS_regularized_energies_phenomenological}),
through the constant factor $\eta$ of Eq.~(\ref{eq:ratio_geometric-sequence}).

\item
{\em Renormalized scaling\/}: 
the base value $E_{0}$ is an {\em adjustable parameter\/}.

\item
{\em Boundedness from below\/}: 
due to the presence of additional ultraviolet physics.

\end{enumerate}
Correspondingly, the quantity 
$ \kappa_{n'}/ \kappa_{n} =  e^{ -  \pi  (n'-n )/\Theta } $
provides the corresponding inverse
ratio of spatial sizes of the associated wave 
functions, as a more detailed analysis of Eq.~(\ref{eq:ISP_BS_wf_>}) shows.
In a similar manner, Eq.~(\ref{eq:reduced_Schrodinger_u}) can be used 
to analyze the scattering sector of the theory, in which the S matrix 
also reproduces 
the spectrum~(\ref{eq:cutoff_BS_regularized_energies_phenomenological})
from its pole structure.

The most remarkable property of the renormalized conformal system is its
attendant geometric scaling. In essence, it describes the 
{\em residual discrete scale invariance\/} (under a discrete subgroup of 
scalings),
which is left over after the 
symmetry breaking inherent in the renormalization process:
\begin{equation}
\epsilon_{n'+m,n+m} =\epsilon_{n',n} 
\label{eq:discrete_scale_invariance}
\end{equation}
(for $m \in Z$) and states that
$E_{n'} /E_{n} 
= E_{n'-n}/E_{0} $.
Reciprocally, the remnant symmetry~(\ref{eq:discrete_scale_invariance}) 
completely characterizes the spectrum: 
its iterative use as a recursion relation implies
a geometric bound-state tower with $E_{n} = E_{0} \, \eta^{n}$,
in which the scaling factor~(\ref{eq:ratio_geometric-sequence})
is determined by comparison with the conformal interaction.
As a result, the spectrum looks identical from any ``vantage point,''
provided that an appropriate proportional rescaling is simultaneously 
enforced. Mathematically, the spectrum
is invariant under all discrete magnifications $\eta^{q}= E_{2}/E_{1}$
(with $q \in Z$) accompanied by a simultaneous shift of ``vantage point'':
$E_{1} \rightarrow E_{2}$.

In practice, the conformal tower is usually
limited in the infrared and experimental
detection of multiple bound states may prove difficult to achieve.
The only crucial requirement for the
applicability of the effective framework is
the existence of a conformally-invariant
domain that sets in around 
an ultraviolet scale $L_{\rm UV}\sim a$, and possibly
limited by an infrared cutoff $L_{\rm IR} \gg L_{\rm UV}$.
In particular, when 
the phenomenological parameters  
 $L_{\rm UV}$ and $L_{\rm IR}$ are  finite,
the number $N_{\rm conf}$
of conformal bound states 
is also finite and can be derived with
 $n \sim N_{\rm conf}$ as an ordinal number, through
inversion of
Eq.~(\ref{eq:cutoff_BS_regularized_energies_phenomenological}); thus,
\begin{equation}
N_{\rm conf} 
\sim \frac{\Theta}{\pi} 
\,
\ln \left( \frac{L_{\rm IR}}{L_{UV} } \right) 
\;  ,
\label{eq:number_conformal_states}
\end{equation}
which is in agreement with standard upper bounds for
the number of bound states~\cite{Calogero}.
Moreover, such bounds enhance the predictability of the conformal
approach by quantifying corrections
associated with the existence of an infrared cutoff,
as shown in Ref.~\cite{EFT_CQM}---where these techniques
are applied to the formation of molecular dipole-bound anions.

\section{\large\bf I\lowercase{ntrinsic 
and core renormalization frameworks}}

Next, we summarize 
the implementation of alternative 
renormalization frameworks in which the limit
$\xi = \kappa a \rightarrow 0$
is applied under the following conditions:

(a) $\kappa$ is to be fixed
by the finite value of the corresponding bound-state energy.

(b)
One of the coupling parameters 
(either $\lambda$ or $\aleph$)
should accordingly run with respect to $a$,
to guarantee that Eq.~(\ref{eq:log_derivatives_asymptotic}) be satisfied.

In the 
\textbf{\textit{intrinsic renormalization framework\/}},
the {\em conformal coupling  $\lambda$ 
is promoted to a running parameter\/}, so that 
$ \Theta = \Theta (a)$, while 
the strength $\aleph$ of the regularizing core interaction
is kept constant.
When this limit is enforced,
the expression $
{\cal L}^{(<)} \left( {\aleph} \right)
/\Theta
$
on the right-hand side of Eq.~(\ref{eq:log_derivatives_asymptotic})
takes a definite value; thus, the corresponding
left-hand side yields a particular value of the function
$
\alpha 
\left( 
\Theta, \kappa a 
\right)
$;
 by consistency with
Eq.~(\ref{eq:alpha_definition}), this
implies the condition
 \begin{equation}
\Theta (a)
\stackrel{(  a \rightarrow 0)}{\sim} 
0
\; \; \; 
{\rm or}
\; \; \; 
\lambda (a)
\stackrel{(  a \rightarrow 0)}{\sim} 
\lambda^{(*)}
 \; 
\label{eq:running_conformal_parameter}
\end{equation}
for the running of the coupling
towards its critical value.
Furthermore, 
Eq.~(\ref{eq:ratio_geometric-sequence})
implies the {\em formal infrared
collapse of the bound-state spectrum\/}:
if the ground-state energy is fixed, the 
conformal tower of excited states is pushed towards its
{\em accumulation point $E=0$\/}; however,
an effective reinterpretation should still lead to the familiar 
sequence~(\ref{eq:cutoff_BS_regularized_energies_phenomenological}).

In addition to the running 
coupling~(\ref{eq:running_conformal_parameter}),
an {\em effective boundary condition\/} can be derived from
Eq.~(\ref{eq:pythagorean}),
if one assumes the continuous matching 
\begin{displaymath}
\left.
V^{(<)}({\bf r})
\right|_{r=a}  = 
\left.
V^{(>)}({\bf r})
\right|_{r=a}    
\; .
\end{displaymath}
Then, as $a \rightarrow 0$ and
defining the variables
$\tilde{\xi} = \tilde{k} a$
and
$\xi = \kappa a$,
with fixed $\kappa$,
the nature of the limit
\begin{equation}
\tilde{\xi} \equiv
 \tilde{k} a
= 
\sqrt{ \aleph - \xi^{2} }
\stackrel{( a \rightarrow 0)}{\sim} 
\sqrt{ \aleph } 
\geq 
\sqrt{ \lambda^{(*)} + \Theta^{2} } 
\; 
\label{eq:xi_tilde_def}
\end{equation}
 depends on whether
$
\lambda^{(*)}
$ 
vanishes or not.
Here, the matching condition
$ \aleph
\left. {\mathcal F} ({\bf r})
\right|_{r=a}= \lambda$
was used, 
so that
$\aleph \geq \lambda$.
The two ensuing scenarios 
($\lambda^{(*)} \neq 0$ 
and 
$\lambda^{(*)} = 0$)
are discussed next.

For $d\neq 2$ or $l \neq 0$: 
$
\lambda^{(*)}
 \neq 0 $
and Eq.~(\ref{eq:xi_tilde_def}) gives 
\begin{displaymath}
\tilde{\xi} 
\stackrel{( a \rightarrow 0)}{\sim} 
\sqrt{ \aleph } 
\geq 
\sqrt{ \lambda^{(*)} } 
> 0
\; ;
\end{displaymath}
this implies that
$
{\cal L}^{(<)} (\tilde{\xi}; \tilde{k} )
  \stackrel{(a \rightarrow 0)}{\sim}
{\cal L}^{(<)} \left( {\aleph} \right)
$
takes a finite and nonzero value---for example,
for a constant
regularizing core~\cite{landau:77},
the reduced wave function
$ 
w_{l+\nu}( \tilde{k} r; \tilde{k})
\propto J_{l+\nu} (\tilde{k} r)$
satisfies this condition, as 
$J'_{l+\nu}(l+\nu) \neq 0$.
Correspondingly, in Eq.~(\ref{eq:log_derivatives_asymptotic}),
$| \cot \alpha | \rightarrow \infty$, so that
\begin{eqnarray}
\sin \alpha (\Theta, \kappa a)
&  \stackrel{(a \rightarrow 0)}{\sim} &
0
\label{eq:pseudo_Dirichlet_sine}
\;  ,
\\
u (r =a)
&  \stackrel{(a \rightarrow 0)}{\sim} &
0
\; ,
\label{eq:pseudo_Dirichlet}
\end{eqnarray}
with
$u(r) = \sqrt{r} \,
v(r)$
being the usual reduced wave function
(cf.\ Eq.~(\ref{reduced_Bessel-like_radial_wf})).
Thus, this framework is compatible with the choice of a 
Dirichlet boundary condition at the shifted position $r=a$.
In fact, Eq.~(\ref{eq:pseudo_Dirichlet}) provides the starting point for the
alternative renormalization framework advanced in
Refs.~\cite{gup:93,camblong:isp}---and also
used in path integral treatments of conformal quantum 
mechanics~\cite{pi_collective}.
However, the more general approach presented in this Letter
sheds light on the emergence of
this {\em effective boundary condition\/}.

For $d=2$ and $l=0$:
$\lambda^{(*)} = 0$
and
\begin{displaymath}
\tilde{\xi} = \tilde{k} a 
\stackrel{( a \rightarrow 0)}{\sim} 
 \sqrt{ \aleph } 
\geq
 \sqrt{ \lambda} 
= O(\Theta)
\; ,
\end{displaymath}
which suggests that
Eq.~(\ref{eq:log_derivatives_asymptotic}) acquires a different limit,
$\cot \alpha \rightarrow 0$, at least for a constant core.
The {\em effective boundary condition\/} becomes
\begin{equation}
\cos \alpha (\Theta, \kappa a)
  \stackrel{(a \rightarrow 0)}{\sim}
0
\;  .
\label{eq:pseudo_Neumann_cosine}
\end{equation}
Even though this has the
appearance of a Neumann boundary condition,
Eq.~(\ref{eq:pseudo_Dirichlet})
is still satisfied
due to the prefactor $\sqrt{r}$.

The intrinsic framework is reminiscent of the 
renormalization theory that predated the effective field theory
program~\cite{EFT}, with a running coupling 
leading to the renormalization-group $\beta$ 
function, defined from
$
\beta (\Theta)
\equiv
\Lambda \,
\partial \Theta  (\Lambda)/\partial \Lambda
$ (with $\Lambda = 1/a$), 
so that
\begin{equation}
\beta (\Theta)
  \stackrel{(a \rightarrow 0)}{\sim}
- \frac{1}{\pi f_{n} }
\,
\Theta^{2}
\; ,
\label{eq:beta_ISP_reg}
\end{equation}
which should be limited to the ground state ($n=0$).
Once the general properties of this framework are understood through
a unified real-space approach, one can consider alternative regularization
schemes within this traditional paradigm
and further examine the running behavior of the coupling constant
(as in the QED realizations of Section~\ref{sec:physical_realizations}).

Finally, in the 
\textbf{\textit{core renormalization framework\/}},
the strength $\aleph$ of the regularizing core
is promoted to a running coupling $ \aleph = \aleph (a) $,
while  the {\em conformal coupling $\lambda$ 
remains fixed\/}~\cite{beane:00,bawin:03}.
In this framework, 
our unified description leads to
\begin{equation}
{\cal L}^{(<)} \left( \aleph (a) \right)
 \stackrel{(a \rightarrow 0)}{\sim} 
\varpi (a) \equiv \Theta 
\cot \left[
\alpha 
\left( 
\Theta, \kappa a 
\right)
\right]
\;  ,
\label{eq:core_renormalization}
\end{equation}
while the rapidly oscillating {\em conformal\/} behavior
of Eq.~(\ref{eq:alpha_definition})
provides the characteristic log-periodic
running coupling $ \aleph (a)$ of the core---the 
celebrated {\em limit cycle\/}
for the renormalization group of the three-body 
problem~\cite{beane:00,bawin:03,3_body_nucleon}.
For $d=3$, $l=0$, and zero energy, 
\begin{equation}
{\cal L}^{(<)} \left( \aleph (a) \right)
=
\sqrt{\aleph (a) }
\,
\cot \sqrt{\aleph (a) } - \frac{1}{2}
\; ,
\label{eq:limit_cycle_d=3_l=0_E=0}
\end{equation}
whose form is also valid for $E \neq 0$ with the replacement 
$\sqrt{\aleph} 
\rightarrow \tilde{k} a $.
In the three-body problem,
a delta-function counterterm is usually modeled
with a square well:
${\mathcal F} ({\bf r})= {\rm const}$ (for $0 \leq r <a$).
Most importantly, our derivation 
is {\em robust\/} and completely general---valid 
for any core ${\mathcal V} ({\bf r})$,
dimensionality, and angular momentum.
For example,
for a flat core in $d \neq 3$ or $l \neq 0$,
the function~(\ref{eq:limit_cycle_d=3_l=0_E=0})  becomes
\begin{displaymath}
{\cal L}^{(<)} \left( \aleph (a) \right)
=
\left. 
\frac{
\tilde{\xi} J_{l+ \nu}'( \tilde{\xi})}{
J_{l+\nu} (\tilde{\xi}) }
\right|_{\tilde{\xi} = \tilde{k} a }
\; .
\end{displaymath}

In short, the main and critical difference between the intrinsic
and core frameworks consists in the {\em treatment of the conformal
coupling\/}:
\begin{itemize}
\item
For a running
conformal coupling dictated by self-consistency requirements
of the conformal interaction:
the intrinsic framework applies.
\item
For a fixed conformal coupling dictated by the long-distance physics:
the core framework is mandatory.
\end{itemize}
In the intrinsic framework, a single
symmetry-breaking bound states survives, as discussed in 
Refs.~\cite{camblong:dt,camblong:isp}.
By contrast, from
the treatment of the conformal coupling as a fixed variable,
{\em 
the conformal physics---including symmetry breaking---of 
the core framework reduces to that of the effective framework 
and displays the\/} 
\textbf{\textit{residual symmetry\/}},
{\em of the strong-coupling regime and its associated geometric scaling.\/}

\section{\large\bf P\lowercase{hysical realizations}}
\label{sec:physical_realizations}

The renormalization frameworks presented in this Letter
provide a unified approach for a broad set of physical realizations. 

\subsection{\large\it 
Near-horizon physics and the thermodynamics of
black holes}

The primary properties of black hole thermodynamics
can be derived within a semiclassical approach in which the quantum fields 
encode the quantum properties
of a black hole~\cite{thooft:85}.
The ensuing physics in generalized Schwarzschild coordinates
is conformally invariant near the
horizon~\cite{near_horizon1,near_horizon2};
this can be shown by considering
a scalar field $\Phi$ described by the Lagrangian
\begin{displaymath}
 \mathcal{ L} =
-
\frac{1}{2}
\,
\left[
g^{\mu \nu}
\,
\nabla_{\mu} \Phi
\,
\nabla_{\nu} \Phi
+
m^{2} \Phi^{2}
+  \xi R \Phi^{2}
\right]
\end{displaymath}
(with mass $m$ and coupling  $\xi$
to the curvature scalar $R$)
in a class of metrics
in $D$ spacetime dimensions,
\begin{displaymath}
ds^{2}
=
- f (r) \,  dt^{2}
+
\left[ f(r) \right]^{-1} \, dr^{2}
+ r^{2} \,
 d \Omega^{2}_{(D-2)}
\; ,
\end{displaymath}
including the Schwarzschild and 
Reissner-Nordstr\"{o}m geometries~\cite{BH_CQM}.
In this approach, a hierarchical expansion
in the radial variable $x=r-r_{+}$,
away from the outer horizon  $r_{+}$, leads to a 
 {\em one-dimensional near-horizon  strong-coupling  conformal interaction\/}
$V(x) \sim - \lambda/x^{2}$, in which $\lambda = \Theta^{2}+ 1/4$
with {\em conformal parameter\/} 
\begin{equation}
\Theta= \frac{\omega}{f'_{+} }
\end{equation}
(for a frequency component $\omega$) and $f'_{+}= f'( r_{+})$.
The corresponding density of states---an appropriate
generalization of the conformal density of states derived from
Eq.~(\ref{eq:number_conformal_states})---experiences an 
{\em ``ultraviolet catastrophe''\/} and
requires a {\em geometric renormalization\/}
equivalent to an effective ``brick wall'' near the 
Planck scale~\cite{thooft:85}.
In essence, this procedure amounts to the presence
of an ultraviolet cutoff $L_{\rm UV}$
for the conformal potential
(with a scale of the order of
$r_{+}$ leading to an infrared cutoff $L_{\rm IR}$). 
Thus, the existence of a conformal domain completely 
drives the thermodynamics, with
the Hawking temperature $T_{H}$ 
uniquely given by the conformal parameter,
\begin{equation}
T_{H}=
\frac{  f'_{+}}{4 \pi} =\frac{ \omega}{4 \pi \Theta} 
\; ;
\label{eq:Hawking_temperature}
\end{equation}
this suggests a conformal derivation of the Hawking effect,
as we  will discuss elsewhere.

\subsection{\large\it Molecular dipole-bound anions}

These systems, formed by the interaction of an electron with a 
polar molecule, possess a conformally invariant domain, with 
an anisotropic generalization of Eq.~(\ref{eq:external_ISP}):
\begin{displaymath}
{\mathcal V^{(>)}} ({\bf r}) = -\frac{\lambda \, \cos \theta}{ r^{2} }
\; .
\end{displaymath} 
Here, a large body of
experimental and computational evidence shows the {\em existence of
a critical dipole moment\/}, in agreement with the conformal 
prediction~\cite{molecular_dipole_anomaly}.
The effective {\em conformal parameter\/} 
\begin{equation}
\Theta
= \sqrt{\gamma -\frac{1}{4}}
\end{equation}
 is related to the dimensionless 
molecular dipole moment $\lambda$
through the roots
of an angular secular equation $D(\gamma, \lambda) = 0$
involving multiple angular momentum channels:
\begin{displaymath}
D(\gamma, \lambda) ={\rm det} 
\left[ (\hat{L}/\hbar)^{2} + \gamma \openone - \lambda \cos \theta \right]
\end{displaymath}
(with $\hat{L}$ being the angular momentum and $\openone$ the 
identity operator),
 in which the matrix elements are evaluated in the angular momentum basis
$\left| l,m \right\rangle$.
This problem gives
a conformal critical dipole moment
$
\lambda^{(*)}_{\rm conf} \approx 1.279
$
(corresponding to
1.625 D, in debye units)
dictated by the 
conformal critical point $\gamma^{(*)} = 1/4$.
In addition, the conformal behavior is limited by
an ultraviolet boundary $L_{\rm UV} \sim a$ of the
order of the molecular size $a$ and
an infrared scale  $L_{\rm IR}$ due to 
the coupling with rotational molecular degrees of 
freedom~\cite{garrett}:
\begin{displaymath}
L_{\rm IR}= r_{B} \equiv \sqrt{\hbar^{2}/ (2 m_{e}B )}
\gg
a \sim L_{\rm UV}  
\; ,
\end{displaymath}
where $B = \hbar^{2}/2I$ is  the rotator constant,
$I \sim M a^{2}$ is the moment of inertia,
 $m_{e}  $ the electron mass, 
and  $M$ is the mass of the molecule.

Incidentally, it should be noticed that an alternative
viewpoint for molecular dipole-bound anions 
was presented in Ref.~\cite{bawin:04},
within an approach centered on the rotational degrees 
of freedom and governed by an effective adiabatic inverse fourth potential.
This alternative treatment, however, fails to account for the 
existence of a critical dipole moment and otherwise does not add any
new ingredients to the physics of electron binding
by polar molecules.
The incompleteness of the proposal of Ref.~\cite{bawin:04}
is in sharp contrast with the conformal approach:
as shown in Ref.~\cite{EFT_CQM}, the physical origin of criticality
and the physics of the logarithmic corrections to the critical dipole moment
are traced directly to the presence of a conformal
window of scales and its associated renormalization---with the 
rotational dynamics merely setting the infrared scale.
Thus, using the parameters
$L_{\rm UV} $ and $L_{\rm IR}$,
a systematic approximation scheme can be introduced to
account for the effects of the rotational degrees of freedom---for example,
the critical dipole moment 
$\lambda^{(*)}
=
\lambda^{(*)}_{\rm conf}\, 
\left( 1 + \epsilon \right)
$
can be computed from
\begin{displaymath}
\epsilon
\approx  
\frac{
80  \, \pi^{2}
\left( 1 - \delta \right)^{2} }{
9 \left[ \ln \left(  r_{B}/a \right)^{2}
\right]^{2} }
\end{displaymath} 
(in which $\delta$ absorbs ultraviolet and infrared
corrections),
with remarkable numerical accuracy~\cite{EFT_CQM}.

\subsection{\large\it The many-body Efimov effect}

This phenomenon,
which consists in the formation of spatially extended bound states 
in a three-body system~\cite{efimov_effect},
has been recently highlighted for the three-body nucleon 
interaction~\cite{3_body_nucleon}.
Specifically, the internal dynamics of the 
three-body system in three dimensions involves 6 degrees of freedom;
a hyperspherical adiabatic expansion~\cite{hyperspherical_adiabatic}
and a Faddeev decomposition of the wave function~\cite{faddeev_eqs}
lead to the Efimov effect for the ensuing
adiabatic potentials~\cite{jensen_Phys_Rep_Efimov}:
the formation of spatially extended and weakly bound states,
with an accumulation point at zero energy.
This is just the {\em conformal tower\/} of bound states,
which can be established through 
the reduction process leading to a 6-dimensional realization of
the conformal interaction~(\ref{eq:external_ISP});
then, Eq.~(\ref{eq:critical_coupling}) for 
$l=0$ implies that $\lambda^{(*)} = 4$, so that $\lambda = 4 + \Theta^{2}$
and the {\em conformal parameter\/} $\Theta$
only depends upon the three ratios of particle masses
(for large scattering lengths)---which, for 
typical physical parameters,
yields a problem in the strong-coupling regime. 
For example, 
for identical bosons with zero-range two-particle interactions,
the corresponding transcendental equation 
\begin{equation}
8 \,
\sinh \left( \frac{\pi \Theta}{6} \right)
= \sqrt{3} \, \Theta \,
\cosh
\left(
\frac{ \pi \Theta }{ 2}
\right)
\end{equation}
provides the value
$\Theta \approx 1.006$.

Thus, the number $N_{E}$
of Efimov bound states 
is approximately given by
Eq.~(\ref{eq:number_conformal_states}), with
the infrared cutoff
$L_{\rm IR} 
\sim
l_{\rm sc}$
(average two-body scattering length)
and the ultraviolet cutoff
$L_{\rm UV} 
\sim
a \sim R_{e}$
(effective range of the interaction).
The Efimov effect---being a characteristic three-body 
phenomenon---has also been applied to the description of the
atomic helium trimer $^{4}{\rm He}_{3}$
and to other atomic and molecular 
combinations~\cite{jensen_Phys_Rep_Efimov,jensen&fedorov_braaten}.

\subsection{\large\it Quantum electrodynamics (QED)}

 Several regimes of 
QED$_{D}$ (in $D=d+1$ spacetime dimensions)
reduce to conformal quantum mechanics and confirm that
chiral symmetry breaking occurs for sufficiently large couplings.
A typical reduction scheme is based on the linearization of the
Euclidean Schwinger-Dyson equation for the fermion self-energy,
followed by a real-space
reinterpretation in terms of an effective 
Schr\"{o}dinger equation~(\ref{eq:reduced_Schrodinger_u})
with $l=0$, within the ladder approximation~\cite{gusynin1,gusynin2, gusynin3};
the existence of bound states in the effective problem
is equivalent to {\em dynamical chiral symmetry breaking\/}
for QED$_{D}$.
The first relevant {\em conformal regime\/} is 
that of  QED$_{3}$ with $N_{f}$ Dirac-fermion 
flavors~\cite{gusynin1}, for intermediate distances,
with {\em conformal parameter\/}
(in three dimensions)
\begin{displaymath} 
\Theta = \sqrt{ \frac{32}{ 3 \, \pi^{2} \, N_{f} } - \frac{1}{4} }
\; ;
\end{displaymath}
thus, there exists a {\em critical fermion number\/} 
\begin{displaymath}
N^{(*)}  
= \frac{128}{ 3  \pi^{2} }
\approx 4.323
\end{displaymath}
for the appearance of the symmetry-breaking tower of conformal 
states~(\ref{eq:cutoff_BS_regularized_energies_phenomenological})---in 
agreement with Refs.~\cite{gusynin1}
and \cite{nash}.
Another {\em conformal regime\/} is 
that of quenched QED$_{4}$,
with {\em conformal coupling\/} 
$\lambda =3 \alpha/\pi $,
proportional to the QED fine structure
constant $\alpha= e^{2}/(4 \pi \hbar c)$;
the {\em conformal parameter\/} is
\begin{equation}
\Theta = \sqrt{ \frac{ 3 \, \alpha}{ \pi} -1}
\; 
\end{equation}
(as Eq.~(\ref{eq:critical_coupling}) gives $\lambda^{(*)} = 1$
in four dimensions),
which  implies a {\em critical QED$_{4}$ coupling  
$\alpha^{(*)} = \pi/3$\/}
for the occurrence of chiral symmetry
breaking~\cite{gusynin2};
by contrast, in lower dimensionalities, 
QED$_{D}$ does not show criticality:
symmetry breaking always occurs
because a nonconformal attractive
regular potential is involved.
Finally,
the running of the effective conformal coupling
in quenched QED$_{4}$
can also be described within the {\em intrinsic 
renormalization framework\/}
using dimensional regularization~\cite{gusynin2, camblong:dt}:
$(\alpha - \alpha^{(*)}) / \alpha^{(*)}
\propto \epsilon^{2/3} $,
 with $\epsilon = (4-D)/2$.

\section{\large\bf C\lowercase{onclusions}}
\label{sec:conclusions}

Renormalization of a conformally invariant 
interaction is mandatory when the ultraviolet physics of the associated 
singular problem dictates the existence of bound states.
In this Letter, we have introduced a generic
 regularization approach in real space and displayed 
the advantages of the effective renormalization framework. 
In conformal quantum mechanics, this procedure leads to an 
 {\em anomaly or quantum symmetry breaking\/} in the strong-coupling 
regime~\cite{cam:anomaly_qm_prelim,cam:anomaly_qm_ISP_anyD,esteve:anomaly}---a
process that is induced by the need to regularize the theory with
a symmetry-breaking dimensional parameter~\cite{camblong:dt} and is
manifested by anomalous terms in the SO(2,1) 
algebra within all renormalization frameworks~\cite{cam:anomaly_qm_ISP_anyD}.
The central properties of near-horizon black-hole thermodynamics,
the formation of dipole-bound anions, the many-body Efimov effect, and 
various regimes of quantum electrodynamics---among other systems---constitute 
an expanding set of effective realizations of this conformal anomaly.
In closing, we emphasize the generality of the techniques
introduced in this Letter, which could 
also be applied to other singular interactions, to the  Calogero 
model~\cite{gupta:calogero}, and possibly to other instances of conformal 
behavior and dynamical symmetry breaking in gauge theories.

\bigskip

\bigskip

\bigskip

{\bf 
Acknowledgments}

This research was supported
by the National Science Foundation under Grant 
No.\ 0308300
(H.E.C.) and under Grant No.\ 0308435 (C.R.O.), and
by the University of San Francisco Faculty Development Fund
(H.E.C.).
We also thank Dr.\ Stanley Nel for generous travel support;
Professors Cliff Burgess, Luis N. Epele,  Huner Fanchiotti,
and Carlos A. Garc\'{\i}a Canal
for discussions leading to this work; and Dr.
Valery P. Gusynin for bringing important references to our attention.

\end{document}